\PassOptionsToPackage{dvipsnames}{xcolor}

\documentclass[runningheads]{llncs}
\usepackage[T1]{fontenc}
\usepackage{graphicx}
\usepackage[relative,overlay]{textpos}
\usepackage{fancyvrb}
\usepackage{tikz}
\usepackage{booktabs}
\usepackage{pifont}
\usepackage{colortbl}
\usepackage[hyphens]{url}

\newcommand*\circledorange[1]{\tikz[baseline=(char.base)]{
    \node[shape=circle,fill=orange,text=white,inner sep=2pt] (char) {#1};}}

\def\checkmark{\tikz\fill[scale=0.4](0,.35) -- (.25,0) -- (1,.7) -- (.25,.15) -- cycle;}

\newcommand{\xmark}{\ding{55}}

\begin{document}
\title{From GPT-3 to GPT-4: On the Evolving Efficacy of LLMs to Answer Multiple-choice Questions for Programming Classes in Higher Education}
\titlerunning{From GPT-3 to GPT-4: On the Evolving Efficacy of LLMs to Answer MCQs}
%

\author{Jaromir Savelka\orcidID{0000-0002-3674-5456}, Arav Agarwal\orcidID{0000-0001-9848-1663}, Christopher Bogart\orcidID{0000-0001-8581-115X} and Majd Sakr\orcidID{0000-0001-5150-8259}}
\institute{School of Computer Science, Carnegie Mellon University, Pittsburgh, PA, USA\\
\email{jsavelka@cs.cmu.edu}}
\authorrunning{Savelka et al.}
%
%
\maketitle              
\begin{abstract}
We explore the evolving efficacy of three generative pre-trained transformer (GPT) models in generating answers for multiple-choice questions (MCQ) from introductory and intermediate Python programming courses in higher education. We focus on the differences in capabilities of the models prior to the release of ChatGPT (Nov '22), at the time of the release, and today (i.e., Aug '23). Recent studies have established that the abilities of the OpenAI's GPT models to handle assessments originally designed for humans keep increasing as the newer more capable models are released. However, the qualitative differences in the capabilities and limitations of these models to reason about and/or analyze programming MCQs have been under-explored. We evaluated three OpenAI's GPT models on formative and summative MCQ assessments from three Python courses (530 questions) focusing on the qualitative differences in the evolving efficacy of the subsequent models. This study provides further evidence and insight into the trajectory of the current developments where there already exists a technology that can be utilized by students to collect passing scores, with no effort whatsoever, on what today counts as viable programming knowledge and skills assessments. This study could be leveraged by educators and institutions to better understand the recent technological developments in order to adapt the design of  programming assessments as well as to fuel the necessary discussions into how assessments in future programming classes should be updated.

\end{abstract}
\section{Introduction}
\label{sec:introduction}
The unveiling of OpenAI's ChatGPT\footnote{ChatGPT. \url{https://chat.openai.com/} [Accessed 2023-01-26]} has sparked significant public discourse about the implications of GPT models in the educational realm. In response to potential risks, such as facilitating plagiarism or dispensing erroneous or unsuitable content, New York City's public school system has barred its use~\cite{ElsenRooney2023}. At the same time, higher education institutions are re-calibrating their assignments~\cite{Huang2023} and are turning to AI-generated text detectors like GPTZero~\cite{Bowman2023}. This paper reinforces and builds upon prior investigations \cite{savelka2023thrilled}, indicating that programming educators should anticipate a future where students can effortlessly use readily available technology to earn passing grades from current programming knowledge and skills evaluations.

Specifically, this paper analyzes the evolving capabilities of generative pre-trained transformers (GPT) to pass typical assessments, i.e., multiple-choice question (MCQ) tests, in introductory and intermediate programming courses at the higher education level. The aim of this paper is to provide further insight into the differences among the successive generations of GPT models, complementing the earlier studies which focus mostly on benchmarking the models \cite{Savelka2023,savelka2023large,savelka2023thrilled}. Here, we investigate qualitative differences in capabilities of the models prior to the release of ChatGPT (Nov'22), at the time of the release, and today~(i.e., Aug '23). To support our experiments we use the same data set as the earlier studies, comprising of 530 MCQ assessments from three Python courses. In prior work, we organized the questions into meaningful categories according to their type (e.g., true/false questions, or questions asking about an output of the provided code snippet). Here, we perform a thematic analysis to identify systematic differences among the studied GPT models in their programming MCQ answering capabilities.

To investigate the evolving capabilities of the three generations of GPT models to handle the MCQ assessments in a programming education context, we analyzed the following research questions:

\begin{enumerate}
    \item What are the systematic differences between OpenAI's \verb|text-davinci-001| (GPT-3) and \verb|text-davinci-003| (GPT-3.5) in how they handle programming MCQs.~(RQ1)
    \item What are the differences between \verb|text-davinci-003| (GPT-3) and \verb|gpt-4| (GPT-4) in handling the MCQs. (RQ2)
\end{enumerate}

\noindent By carrying out this work, we provide the following contribution to the computing education research community. To the best of our knowledge, this is the first comprehensive study that performs systematic qualitative evaluation via thematic analysis of the differences in performance on MCQ-style assessments, often involving code snippets, of different generations of the OpenAI's GPT models.

\section{Background}
\subsection{Motivation}
\label{sec:background}
As a motivation for this study, consider the following Python script that expects a number as the input: 

\begin{quote}
\begin{Verbatim}
try:
    value = input("Enter a value: ")
    print(int(value) / len(value))
except ZeroDivisionError:
    print("Very bad input...")
\end{Verbatim}
\end{quote}

\noindent The user enters a value (type \verb|str|) which is cast to an \verb|int|. The entered number is then divided by the length, i.e., the number of characters, of the raw \verb|str| input. Notably, the code anticipates the possibility of the \verb|ZeroDivisionError|. However, this particular error cannot occur (explained below). 

\begin{figure}[t]
\includegraphics[width=\textwidth]{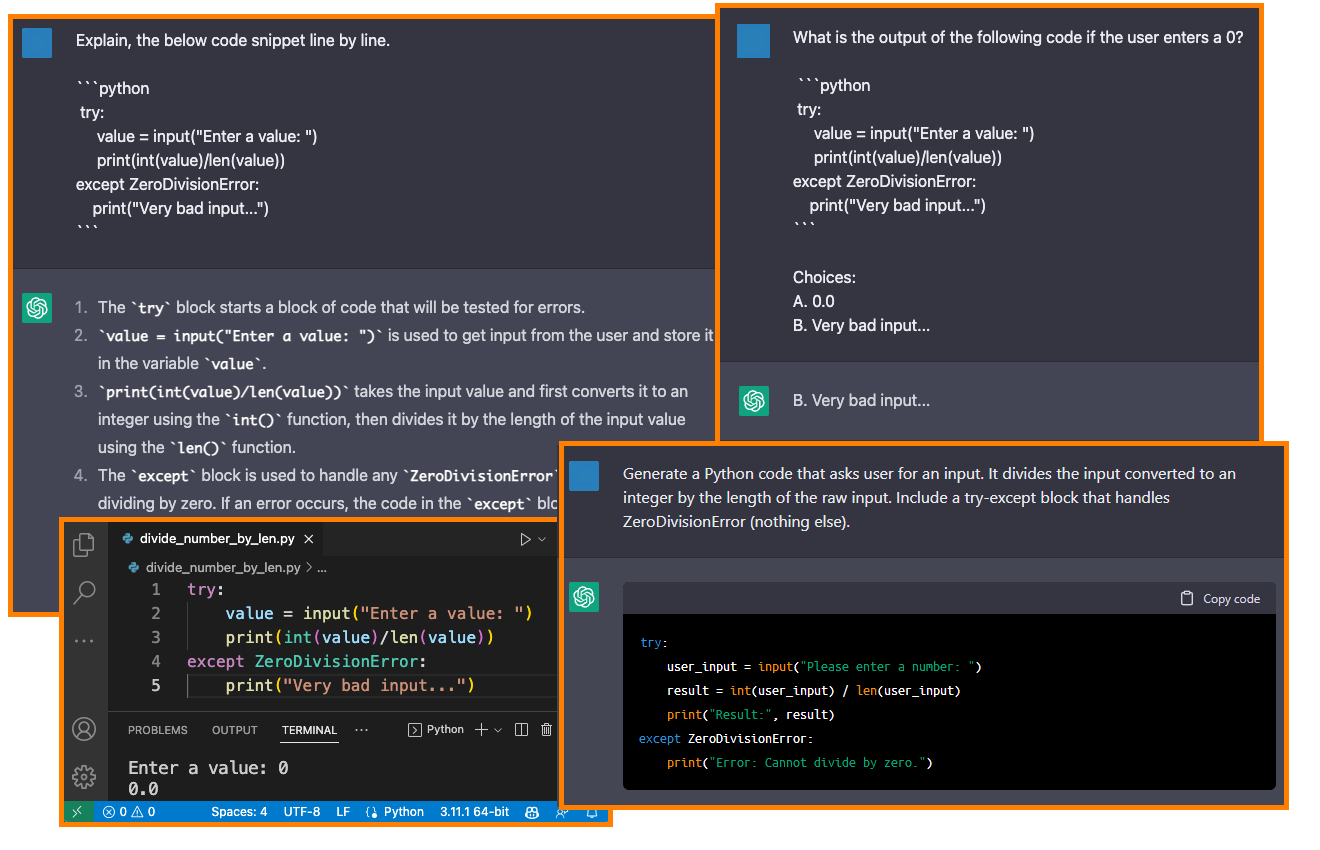}
\begin{textblock*}{3.4cm}(.2cm,-4.3cm)
\circledorange{1}
\end{textblock*}
\begin{textblock*}{3.4cm}(9.15cm,-1.65cm)
\circledorange{2}
\end{textblock*}
\begin{textblock*}{3.4cm}(8.75cm,-4.4cm)
\circledorange{3}
\end{textblock*}
\begin{textblock*}{3.4cm}(2.8cm,-1.00cm)
\circledorange{4}
\end{textblock*}
\caption{The upper-left screenshot depicts a conversation with GPT-3.5 (ChatGPT interface) when asked to explain a code snippet line by line. It correctly explains the behavior (1). The lower-right shows a conversation with the model when asked to generate the code snippet with the same behavior. The generated code is correct (2). The upper-right screenshot depicts a conversation with GPT-3.5 when asked a straightforward MCQ about a code it can correctly explain line by line as well as correctly generate. The answer is wrong (3)---compare the actual output of the code snippet which is shown in the lower-left corner (4). The figure was originally published in \cite{savelka2023large}.}
\label{fig:motivation}
\end{figure}

If a user enters 44 the output would be 22.0 (i.e., \verb|44 / 2| cast into \verb|float| as a result of applying the \verb|/| operator). To get an anecdotal insight into the differences between \verb|text-davinci-001| (GPT-3), \verb|text-davinci-003| (GPT-3.5) and GPT-4 (see Section \ref{sec:models} for additional details on these models), we submitted the following queries to each model related to the provided code snippet:

\begin{enumerate}
    \item Explain the code snippet line by line.
    \item What is the output of the code snippet if the user enters 0.
    \item Generate the code snippet based on the provided description in natural language (see Figure \ref{fig:motivation}).
\end{enumerate}

\noindent The GPT-3 model struggles to explain the code. While it is providing essentially correct statements it operates at a very general (superficial) level: 

\begin{quote}
Python's try statement allows you to execute a block of code, and provides a way to handle errors that may occur.
\end{quote}

\noindent When provided with the code snippet and asked about the output when the user enters 0 (letting the model choose from ``A. 0.0'' or ``B. Very bad input...''), the generated answer is ``B. Very bad input...'' Note that this is an incorrect answer as the length of the string \verb|"0"| is 1. Therefore, the correct answer is ``A. 0.0''~(as shown in Figure \ref{fig:motivation}). A student selecting the incorrect answer could rightly be suspected of having multiple misconceptions. Opting for the ``B.~Very bad input...'' choice could be more understandable if the \verb|value| variable were not placed within the \verb|len()| function call. In such a case, one could simply assume that the student did not recognize that the output of the \verb|input()| function call is a \verb|str|, assuming it was an \verb|int|. However, applying the \verb|len()| function to an \verb|int| raises the \verb|TypeError|. That is why the only input that could potentially result in the \verb|ZeroDivisionError| would be an empty string. Interestingly, even that input would not yield the particular error as it would not succeed in casting the \verb|value| variable to \verb|int| (\verb|ValueError|). Finally, when asked to generate the code snippet based on the provided description the GPT-3 model fails as well, generating the following:

\begin{quote}
\begin{Verbatim}
try: 
    input = int(raw_input)
except ZeroDivisionError: 
    print("Invalid input") 
else: 
    print("The result is: {}".format(input / length)) 
\end{Verbatim}
\end{quote}

\noindent While the code is not completely unlike the expected solution it uses the previously undefined \verb|length| variable in the last line. Hence, it would crash.

Possibly the most intriguing are the completions to the above described prompts generated by the GPT-3.5 model. As shown in Figure \ref{fig:motivation}, this model still answers the MCQ about the output of the code snippet incorrectly. However, when asked to explain the code snippet line by line the model produces a detailed and correct explanation, including the line where the division takes place. For that line, it states that: 

\begin{quote}
[it] takes the input value and first converts it to an integer using the \verb|int()| function, then divides it by the length of the input value using the \verb|len()| function.    
\end{quote}

\noindent As seen in Figure \ref{fig:motivation}, the output of the model when prompted to generate the Python code with the same functionality from the natural language description, is also correct. Hence, the GPT-3.5 model correctly explains the execution of the code snippet line by line (i.e., at a local level). Furthermore, it generates the equivalent computer program from a natural language description. However, it fails in answering a simple question about the very same program. Presumably, this would be in stark contrast to a typical student. A student who could write the program on their own from the natural language description as well as correctly explain its workings line by line, would certainly be in a position to answer the MCQ correctly.

\begin{table}[t]
\centering
  \caption{The table summarizes the handling of the MCQ revolving around the tricky code snippet. While GPT-3 fails to explain the code line by line, nor implement it based on the natural language explanation, nor answer as simple MCQ about it, the state-of-the art GPT-4 model is capable of performing all three activities successfully.}
  \label{tab:motivation}
  \setlength{\tabcolsep}{12pt}
  \begin{tabular}{lccc}
  \toprule
                  & Line by line & Implementation  &                \\
    Model         & explanation  & from NL    & Correct answer \\
  \midrule
    GPT-3         &\cellcolor{red!10} \color{OrangeRed}\xmark      &\cellcolor{red!10} \color{OrangeRed}\xmark      & \cellcolor{red!10}\color{OrangeRed}\xmark \\
    GPT-3.5       &\cellcolor{green!10}\color{OliveGreen}\checkmark &\cellcolor{green!10} \color{OliveGreen}\checkmark &\cellcolor{red!10}\color{OrangeRed}\xmark \\
    GPT-4         & \cellcolor{green!10}\color{OliveGreen}\checkmark & \cellcolor{green!10}\color{OliveGreen}\checkmark &\cellcolor{green!10} \color{OliveGreen}\checkmark \\
  \bottomrule
  \end{tabular}
\end{table}

The GPT-4 model not only produces the correct line-by-line explanation and implementation of the code snippet, it also answers the MCQ correctly. Hence, we observe a clear progression in the capabilities of the three studied models in handling the MCQ that involves the tricky piece of code (summarized in Table \ref{tab:motivation}). The oldest of the three models demonstrates rather poor capabilities in handling of the code snippet and consequently fails to answer the MCQ correctly. The newer GPT-3.5 model manifests what could be considered quite a robust understanding if performed by a student, yet, it still fails to generate the correct answer. Finally, the state-of-the art GPT-4 model not only demonstrates the understanding but also produces the correct answer. This study aims to investigate the systematic differences among the three models in how successfully they handle MCQs from programming courses at the postsecondary education level.

\subsection{Related Work}
\label{sec:related_work}
In prior work, we evaluated the capability of various GPT models to pass a diverse set of assessment instruments, including MCQs, in the realistic context of full-fledged programming courses~\cite{Savelka2023,savelka2023large,savelka2023thrilled}. We found that the state-of-the-art GPT models are capable of confidently passing the full spectrum of assessments typically involved in a Python programming course. In this paper, we further explore the evolution of these models, focusing on discovery of more fine-grained properties of MCQs that used to be challenging for the GPT models to handle but are not anymore.

To the best of our knowledge, there are no other studies of GPT models' handling of MCQs from programming courses. However, there are works analyzing the performance on MCQ datasets from other areas. Robinson et al. use InstructGPT \cite{ouyang2022training} and Codex to answer questions from the OpenBookQA~\cite{mihaylov2018can}, StoryCloze \cite{mostafazadeh2016corpus}, and RACE-m \cite{lai2017race} datasets focused on multi-hop reasoning, recall, and reading comprehension. They report 77.4-89.2\% accuracy \cite{robinson2022}. Drori and Verma apply Codex to write Python programs to solve 60 computational linear algebra MCQs with 100\% accuracy \cite{https://doi.org/10.48550/arxiv.2111.08171}. 
Hendryks et al. created a dataset of MCQs covering various STEM, humanities and arts topics. GPT-3 performs at levels above 50\% for subjects such as marketing and foreign policy. It scores below 30\% for topics such as formal logic~\cite{hendrycks2022}. Further, they observed that GPT-3 performed  poorly in quantitative subjects, such as elementary mathematics. Lu et al. evaluated GPT models on a large dataset of 21,208 MCQs \cite{pan2022}.


In computing education context, LLMs have been shown to be highly effective at generating code and explanations of the code for entry-level programmers~\cite{Sarsa_2022}. Such explanations have even been observed to out-class student explanations of the same code \cite{leinonen2023comparing}. Denny et al. discovered that well-structured prompts could yield correct solutions to many programming exercises \cite{https://doi.org/10.48550/arxiv.2210.15157}. Piccolo et al. demonstrated that LLMs can perform most entry-level programming tasks in the context of introductory bioinformatics course \cite{piccolo2023many}. Phung et al. introduced a system that harnessed LLMs to provide precision feedback on syntax errors in students' code \cite{Phung2023GeneratingHF}.
Such feedback explanations went far beyond describing the code line-by-line.  MacNeil et al. demonstrated that explanations of generated code can be offered at multiple different levels of abstraction \cite{MacNeill,MacNeil2}. Finally, Doughty et al. proposed an LLM-powered pipeline to generate MCQs for programming classes~\cite{doughty2022comparative,agarwal2023understanding}. They showed that in many respects the generated MCQs are comparable to the human-crafted ones and well aligned with learning objectives, that can also be automatically generated \cite{sridhar2023harnessing}.

In the near future, it is reasonable to expect LLMs to facilitate teacher-student exchanges similar to those that only occur in a classroom, and are invaluable to student learning \cite{tan2023applying}. Liffiton et al. described and analyzed the usage of CodeHelp \cite{liffiton2023codehelp,sheese2023patterns}, an early example of a tool capable of such interactions. They also showed how students submit different types of questions (e.g., aimed at debugging or conceptual understanding), envisioning a system that can provide responses specifically tailored to each question type \cite{savelka2023patterns}.

\section{Dataset}
\label{sec:data_set}
To support the experiments in \cite{Savelka2023}, we collected MCQs from the following three Python programming courses:

\begin{enumerate}
    \item \emph{Python Essentials - Part 1 (Basics)}\footnote{OpenEDG: Python Essentials - Part 1 (Basics). Available at: \url{https://edube.org/study/pe1} [Accessed 2023-01-15]} (\textbf{PE1}) guides a student from a state of complete programming illiteracy to a level of programming knowledge which allows them to design, write, debug, and run programs encoded in the Python language. The course consists of four content units and one completion (summary) test. The units include:
    \begin{enumerate}
        \item introduction to Python and computer programming,
        \item data types variables, basic I/O, operations and basic operators,
        \item boolean values, conditional loops, lists, logical and bitwise operators, and 
        \item functions, tuples, dictionaries, data processing and exceptions.
    \end{enumerate}
    In this course, formative assessments are called quizzes while summative assessments are called tests. The tests determine if learners pass the courses whereas quizzes are meant as practice. The MCQs often include small snippets of code for learners to reason about.
    \item \emph{Python Essentials - Part 2 (Intermediate)} (\textbf{PE2})\footnote{OpenEDG: Python Essentials - Part 2 (Intermediate). Available at: \url{https://edube.org/study/pe2} [Accessed 2023-01-15]} focuses on more advanced aspects of Python programming, including modules, packages, exceptions, file processing, object-oriented programming. Similarly to PE1, the course is organized into four content units and one completion (summary) test. The course units are:
    \begin{enumerate}
        \item modules, packages, and pip,
        \item strings, string and list methods, and exceptions, 
        \item object-oriented programming, and 
        \item miscellaneous. 
    \end{enumerate}
    \item \emph{Practical Programming with Python}\footnote{Sail(): Social and Interactive Learning Platform. Available at: \url{https://sailplatform.org/courses}. [Accessed 2023-03-03]} (\textbf{PPP}) emphasizes hands-on experience with fundamental Python constructs and exposure to software development tools, practices, and real-world applications. The course consists of eight units which include:
    \begin{enumerate}
        \item Python basics and introduction to functions, 
        \item control flow, strings, input and output,
        \item Python data structures,
        \item object-oriented programming,
        \item software development,
        \item data manipulation,
        \item web scraping and office document processing, and
        \item data analysis.
    \end{enumerate}
    PPP uses MCQ-style inline activities as formative assessment and tests as summative assessment.
\end{enumerate}

\noindent In this work, we continue to work with this data set. Table~\ref{tab:dataset} has additional details.

\begin{table}[t]
  \centering
  \caption{MCQ Data Set. Each horizontal segment provides information about the MCQ assessments each of the courses employ. Each column reports on the distribution of the MCQ types across the courses. Note: T/F -- True/False, Id. T/F -- Identify True/False Statement, Fin. S. -- Finish Statement, Fill-in -- Fill-in Blanks.}
  \label{tab:dataset}
 \setlength{\tabcolsep}{3.5pt}
  \begin{tabular}{l|r|rrrr|rrrrrr|r}
  \toprule\
    Course        & MCQ Type  & T/F & Id. T/F & Fin. S. & Fill-in & Output & Other & Overall \\
  \hline
    PE1           & no code   & 0   & 5       & 23      & -       & -      & 18    & 46      \\
                  & with code & 0   & 5       & 6       & 0       & 51     & 41    & 103     \\
                  & overall   & 0   & 10      & 29      & 0       & 51     & 59    & 149     \\
  \hline
    PE2           & no code   & 0   & 7       & 31      & -       & -      & 10    & 48       \\ 
                  & with code & 0   & 0       & 21      & 0       & 27     & 52    & 100      \\
                  & overall   & 0   & 7       & 52      & 0       & 27     & 62    & 148      \\
  \hline
    PPP           & no code   & 25  & 32      & 2       & -       & -      & 19    & 78       \\
                  & with code & 23  & 21      & 0       & 13      & 32     & 66    & 155      \\
                  & overall   & 48  & 53      & 2       & 13      & 32     & 85    & 233      \\
  \hline
    Type Overall  &           & 48  & 70      & 83      & 13      & 110    & 206   & 530     \\
  \bottomrule
  \end{tabular}
\end{table}

In \cite{savelka2023large}, simple pattern matching, followed by manual curation, was used to organize the MCQs into several categories. In this work we continue to use the labels produced in this way. The high-level distinction was made between those MCQs \emph{with code} and those MCQs with \emph{no code}. To be deemed as \emph{with code}, one of the following two conditions had to be met for an MCQ:

\begin{enumerate}
    \item Within the body of the question there had to be at least one line fully dedicated to computer code.
    \item The choices were computer code expressions.
\end{enumerate}

\noindent Inline mentions of function names or variables were not sufficient for an MCQ to be considered \emph{with code}.

The more fine-grained distinction was focused on the manner in which the question is expressed. The \emph{True/False} questions asked the student to assess the truthfulness of a statement (correct answer emphasized), e.g.:

\begin{quote}
    Developers that write code individually are not expected to apply code standards.\\
    A. True\\
    \emph{B. False}
\end{quote}

\noindent The \emph{Identify True/False Statement} questions asked the student to pick one or more answer choices that are either true or false. Note that this is different from the \emph{True/False} questions (previous category), e.g.:

\begin{quote}
Take a look at the snippet and choose one of the following statements which is true:
\begin{verbatim}
nums = []
vals = nums[:]
vals.append(1)\end{verbatim}
A. \verb|nums| is longer than `vals`\\
\emph{B.} \verb|vals| \emph{is longer than} \verb|nums|\\
C. \verb|nums| and \verb|vals| are of the same length
\end{quote}

\noindent The \emph{Finish Statement} questions asked the student to complete a statement, e.g.:

\begin{quote}
    Right-sided binding means that the following expression:
   \begin{verbatim}1 ** 2 ** 3\end{verbatim}
    will be evaluated:\\
    \emph{A. from right to left}\\
    B. in random order\\
    C. from left to right
\end{quote}

\noindent The \emph{Output} questions asked the student to identify the choice that corresponds to the output of a given snippet of code. This category is applicable only to questions \emph{with code}, e.g.:

\begin{quote}
What is the output of the following snippet?
\begin{verbatim}
my_list_1 = [1, 2, 3]
my_list_2 = []
for v in my_list_1:
    my_list_2.insert(0, v)
print(my_list_2)\end{verbatim}
A. \verb|[1, 2, 3]|\\
B. \verb|[1, 1, 1]|\\
C. \verb|[3, 3, 3]|\\
\emph{D.} \verb|[3, 2, 1]|
\end{quote}

\noindent The \emph{Fill-in Blanks} questions asked the student to fill in a code snippet by selecting the appropriate choice as an answer. This category is applicable only to questions \emph{with code}, e.g.:

\begin{quote}
Fill in the blank of the \verb|is_negative| function definition shown below, so that the function returns \verb|True| when the argument provided to \verb|num| is a negative number and returns \verb|False| otherwise.
\begin{verbatim}
def is_negative(num):
    return _________________
\end{verbatim}
A. \verb|not (num > 0)|\\
B. \verb|num > 0|\\
C. \verb|num <= 0|\\
\emph{D.} \verb|num < 0|
\end{quote}

\noindent The \emph{Other} questions are any MCQs that do not fall into any of the above categories, e.g.:

\begin{quote}
Given the piece of code presented in the code snippet below, what is the value of \verb|palindromes[1]|?
\begin{verbatim}
palindromes = ['pop', 'noon', 'madam']
\end{verbatim}
A. \verb|'pop'|\\
\emph{B.} \verb|'noon'|\\
C. \verb|'p'|\\
D. \verb|'madam'|\\
E. \verb|'o'|
\end{quote}

\noindent Table \ref{tab:dataset} reports the distribution of the MCQs into the above described categories. 

\section{Experiments}

\subsection{Models}
\label{sec:models}
The original GPT model (i.e., GPT-1) \cite{radford2018improving} is a 12-layer decoder-only transformer~\cite{vaswani2017attention} with masked self-attention heads. Its core capability is fine-tuning on a downstream task. The GPT-2 model~\cite{radford2019language} largely follows the details of the original GPT model with a few modifications, such as layer normalization moved to the input of each sub-block, additional layer-normalization after the first self-attention block, and a modified initialization. Compared to the original model it displays remarkable multi-task learning capabilities~\cite{radford2019language}. The third generation of GPT models (i.e., GPT-3)~\cite{brown2020language} uses almost the same architecture as GPT-2. The only difference is that it alternates dense and locally banded sparse attention patterns in the layers of the transformer. The main focus of Brown et al. was to study the dependence of performance and model size where eight differently sized models were trained (from 125 million to 175 billion parameters). The largest of the models is commonly referred to as GPT-3. The interesting property of these models is that they appear to be very strong zero- and few-shot learners. This ability appears to improve with the increasing size of the model~\cite{brown2020language}. The technical details about the GPT-4 model have not been disclosed due to concerns about potential misuses of the technology as well as highly competitive market with generative AI \cite{openai2023gpt4}.

We focus on the evolving capabilities of the models prior to the release of ChatGPT (Nov '22), at the time of the release, and today (i.e., Aug '23). We employ the  InstructGPT \verb|text-davinci-001| model (GPT-3) to evaluate the performance prior to the ChatGPT's release. To understand the performance close to the date when ChatGPT was released, we use the \verb|text-davinci-003| (GPT-3.5). It builds on top of previous \verb|text-davinci-002|, which in turn is based on \verb|code-davinci-002| (focused on code-completion tasks) which is sometimes referred to as codex.\footnote{OpenAI: Model index for researchers. Available at: \url{https://beta.openai.com/docs/model-index-for-researchers/instructgpt-models} [Accessed 2023-01-15]} As of writing of this paper, GPT-4 is by far the most advanced model released by OpenAI. The model is focused on dialog between a user and a system. Hence, to gauge the rate of improvement over the several recent years, we compare the performance of GPT-4 to GPT-3.5 as well as to the previous generation's GPT-3 on the MCQ answering task.

We set the \verb|temperature| of all the models to 0.0, which corresponds to no randomness. The higher the \verb|temperature| the more creative the output but it can also be less factual. As the temperature approaches 0.0, the model becomes more deterministic, which we deem as important for reproducibility. Given that existing literature does use different temperatures for testing, we did initially test a variety of temperatures, but found that setting temperature to 0.0 worked well for our context. Note that this is quite a common choice in the existing work regarding MCQs  \cite{bommarito2023gpt,Lievin2022CanLL,pan2022}. 
We set \verb|max_tokens| to 500 (a token roughly corresponds to a word). This parameter controls the maximum length of the completion (i.e., the output). Note that each model has a length limit on the prompt, and the completion counts towards that limit. While GPT-4 allows for 8,192 tokens\footnote{There is also a variant of the model that supports up to 32,768 tokens.} the GPT-3.5 can only accept up to 4,097 tokens, and GPT-3 is limited to 2,048. We set \verb|top_p| to 1, as is recommended  when \verb|temperature| is set to 0.0. This parameter is related to \verb|temperature| and also influences creativeness of the output. We set \verb|frequency_penalty| to 0, which allows repetition by ensuring no penalty is applied to repetitions. Finally, we set \verb|presence_penalty| to 0, ensuring no penalty is applied to tokens appearing multiple times in the output.

\subsection{Experimental Design}
\label{sec:experimental_design}
\begin{figure}[t]
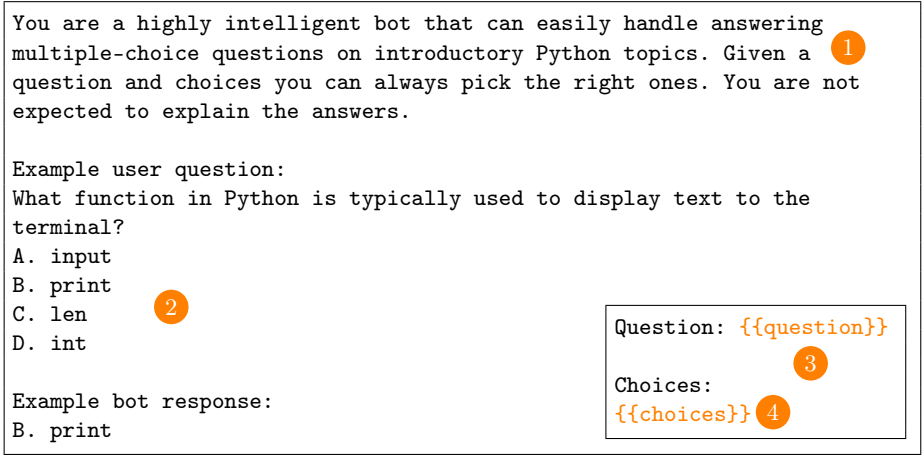

\begin{Verbatim}[frame=single,commandchars=\\\{\}]
You are a highly intelligent bot that can easily handle answering
multiple-choice questions on introductory Python topics. Given a 
question and choices you can always pick the right ones. You are not 
expected to explain the answers.

Example user question:
What function in Python is typically used to display text to the 
terminal?
A. input
B. print
C. len
D. int

Example bot response:
B. print
\end{Verbatim}

\begin{textblock*}{4.0cm}(8.0cm,-2.7cm)
\begin{Verbatim}[frame=single,commandchars=\\\{\}]
Question: \textcolor{orange}{\string{\string{question\string}\string}}

Choices:
\textcolor{orange}{\string{\string{choices\string}\string}}
\end{Verbatim}
\end{textblock*}

\begin{textblock*}{3.4cm}(11cm,-6cm)
\circledorange{1}
\end{textblock*}
\begin{textblock*}{3.4cm}(2cm,-2.55cm)
\circledorange{2}
\end{textblock*}
\begin{textblock*}{3.4cm}(10.5cm,-1.8cm)
\circledorange{3}
\end{textblock*}
\begin{textblock*}{3.4cm}(10cm,-1.15cm)
\circledorange{4}
\end{textblock*}
\caption{MCQ Prompt Templates for GPT-4. The outer frame shows the system's prompt which is used to set the context of the dialogue. The text of the preamble (1) is inspired by OpenAI's QA example. The example user question and bot response (2) primes the model to return the answer(s) only (no explanations). The inner frame depicts the user's message sent to the dialogue system. The \string{\string{question\string}\string} token~(3) is replaced with the question text. The \string{\string{choices\string}\string} token~(4) is replaced with the candidate answers where each one is placed on a single line preceded by a capital letter.}
\label{fig:mcq-prompt-template-gpt4}
\end{figure}

To test the performance of the three GPT models on the task of answering programming MCQs, we submit questions one by one using the \verb|openai| Python library\footnote{GitHub: OpenAI Python Library. Available at: \url{https://github.com/openai/openai-python} [Accessed 2023-08-16]} which is a wrapper for the OpenAI's REST API. For GPT-3 and GPT-3.5, we embed each question in the prompt template shown in Figure \ref{fig:mcq-prompt-template}. Since GPT-4 is a model optimized for dialogue, we use different prompts---the ones shown in Figure \ref{fig:mcq-prompt-template-gpt4}. Note that the prompt for GPT-4 is designed with the intent to prevent the model from explaining the answer to a user as we are only interested in the answer(s) themselves. Each model returns one or more of the choices as the prompt completion (response), which is then compared to the reference answer. Partially correct answers are considered to be incorrect.

\begin{figure}[t]
\footnotesize
\begin{Verbatim}[frame=single,commandchars=\\\{\}]
I am a highly intelligent bot that can easily handle answering 
multiple-choice questions on introductory Python topics. Given 
a question and choices I can always pick the right ones.

Question: \textcolor{orange}{\string{\string{question\string}\string}}

Choices:
\textcolor{orange}{\string{\string{choices\string}\string}}

The correct answer:
\end{Verbatim}
\begin{textblock*}{3.4cm}(10.65cm,-4.00cm)
\circledorange{1}
\end{textblock*}
\begin{textblock*}{3.4cm}(3.95cm,-2.85cm)
\circledorange{2}
\end{textblock*}
\begin{textblock*}{3.4cm}(2.1cm,-1.65cm)
\circledorange{3}
\end{textblock*}
\caption{MCQ Prompt Template for GPT-3 and GPT-3.5. The text of the preamble (1) is inspired by OpenAI's QA example. The 
\{\{question\}\} token~(2) is replaced with the question text. The \{\{choices\}\} token~(3) is replaced with the candidate answers where each one is placed on a single line preceded by a capital letter. The figure was originally published in \cite{savelka2023large}.}
\label{fig:mcq-prompt-template}
\end{figure}

For the quantitative evaluation of the models (Section \ref{sec:results-quant}), we report their performance across the individual units of each course. Furthermore, we show how the three GPT models perform with respect to the MCQ types described in Section \ref{sec:data_set}. Finally, we analyze the differences in the success rates of answering the MCQs across the types between GPT-3 and GPT-3.5 as well as between GPT-3.5 and GPT-4.

For the qualitative analysis of the differences between the successive generations of the GPT models (Section \ref{sec:results-qual}), we conducted a thematic analysis across the subsets of the MCQs to which the preceding model generated a wrong answer whereas the more recent model answered correctly. During the analysis we extracted interesting features of each data point as codes. These were then collated into higher-level themes \cite{braun2006using}. This enabled us to understand the main qualitative improvements between the two studied pairs of the models, i.e., GPT-3 vs GPT-3.5 and GPT-3.5 vs GPT-4.

\section{Results}
\label{sec:results}

\subsection{Quantitative Analysis}
\label{sec:results-quant}
Table \ref{tab:course_results} reports the results of applying the three GPT models to the MCQs from the three courses. These experiments were originally conducted in \cite{Savelka2023,savelka2023large,savelka2023thrilled}. Overall, the GPT-3 model correctly answered 199 out of the 530 questions (37.5\%). GPT-3.5 was more successful with 341 correctly answered MCQs (64.3\%). The most capable GPT-4 model successfully answered 446 questions (84.1\%). Therefore, we conclude that there are noticeable improvements across the successive generations of the GPT models.

\begin{table}[t]
\centering
  \caption{Course results. The table shows the performance of the three studied models from the perspective of the three courses included in the dataset.}
  \label{tab:course_results}
  \setlength{\tabcolsep}{4pt}
  \begin{tabular}{l|rrr}
    \toprule
    Course                            &\multicolumn{1}{|c}{GPT-3}&\multicolumn{1}{c}{GPT-3.5}&\multicolumn{1}{c}{GPT-4}\\
    \midrule
    Python Essentials 1       & 55/149 (36.9\%) & 96/149 (64.4\%)  & 130/149 (87.2\%)  \\
    Python Essentials 2       & 58/148 (39.2\%) & 101/148 (68.2\%) & 134/148 (90.5\%) \\
    Practical Prog. w/ Python & 86/233 (36.9\%) & 144/233 (61.8\%) & 184/233 (79.0\%)\\
    \midrule
    \bf Total                 & \bf 199/530     & \bf 341/530      & \bf 446/530 \\
                              & \bf (37.5\%)    & \bf (64.3\%)     & \bf (84.1\%) \\
    \bottomrule
  \end{tabular}
\end{table}

Table \ref{tab:qtype_results} summarizes how the three GPT models perform on MCQs of various types introduced in Section \ref{sec:data_set}. For all the three models, there is a noticeable difference between their performance on the MCQs \emph{with code} compared to the \emph{no code} MCQs. This is not surprising because the code and natural language combined together are certainly more challenging than natural language alone. It is also possible that the questions containing code snippets may tend to be more difficult than questions that do not. Interestingly, the difference seems to be much more pronounced in case of the GPT-3 (29.9\% vs 53.3\%) and GPT-3.5 (59.5\% vs 77.9\%) models as compared to GPT-4 (81.3\% vs 91.3\%). Therefore, we tentatively conclude that GPT-4 is much more capable in handling MCQs \emph{with code} when compared to GPT-3 and GPT-3.5.

\begin{table}[t]
  \caption{Performance of the GPT models across MCQs of different types.}
  \label{tab:qtype_results}
  \centering
  \begin{tabular}{l|rrr|rrr}
  \toprule
                        &\multicolumn{3}{c}{\bf No Code}    &\multicolumn{3}{|c}{\bf With Code}      \\
  \midrule
    Question Type       &\multicolumn{1}{c}{GPT-3}&\multicolumn{1}{c}{GPT-3.5}&\multicolumn{1}{c}{GPT-4}&\multicolumn{1}{|c}{GPT-3}&\multicolumn{1}{c}{GPT-3.5}&\multicolumn{1}{c}{GPT-4}\\
  \midrule
    True/False          & 13/25      & 20/25      & 23/25   & 12/23      & 10/23      & 13/23 \\
                        & (52.0\%)   & (80.0\%)   & (92.0\%)& (52.2\%)   & (43.5\%)   & (56.5\%)\\
    Identify True/False & 12/44      & 27/44      & 35/44   & 10/26      & 11/26      & 16/26 \\
                        & (27.3\%)   & (61.4\%)   & (79.5\%)& (38.5\%)   & (42.3\%)   & (61.5\%)\\
    Finish Statement    & 42/56      & 50/56      & 56/56   & 10/27      & 22/27      & 25/27\\
                        & (75.0\%)   & (89.3\%)   & (100\%) & (37.0\%)   & (81.5\%)   & (92.6\%)\\
    Fill-in             &-           &-           &-        & 5/13       & 11/13      & 12/13\\
                        &-           &-           &-        & (38.5\%)   & (84.6\%)   & (92.3\%)\\      
    Output              &-           &-           &-        & 28/110     & 53/110     & 86/110\\
                        &-           &-           &-        & (25.4\%)   & (48.2\%)   & (78.2\%)\\
    Other               & 25/47      & 37/47      & 43/47   & 42/159     & 106/159    & 139/159\\
                        & (53.2\%)   & (78.7\%)   & (91.5\%)& (26.4\%)   & (66.7\%)   & (87.4\%)\\
    \midrule
    \bf Total           &\bf 92/172  &\bf 134/172 &\bf 157/172&\bf 107/358 &\bf 213/358 &\bf 291/358   \\
                        &\bf (53.5\%)&\bf (77.9\%)&\bf (91.3\%)&\bf (29.9\%)&\bf (59.5\%)&\bf (81.3\%)\\
  \bottomrule
  \end{tabular}
\end{table}

Table \ref{tab:gpt3_diff} focuses on the differences between the GPT-3 and GPT-3.5 models. Overall, GPT-3.5 fixed 192 (58\%)  incorrect answers provided by GPT-3. On the other hand, GPT-3.5 incorrectly answered 44 questions that were answered correctly by the older GPT-3 model. The rate of improvement appears to be higher for the \emph{no code} MCQs (65\% of mistakes fixed) compared to the \emph{with code} questions (55.7\%). The \emph{no code} questions together with the \emph{other} questions \emph{with code} are the main contributors towards the noticeably higher performance of GPT-3.5 compared to GPT-3. For these MCQs, the GPT-3.5 model corrected 126 questions incorrectly answered by GPT-3 while only committing 20 novel mistakes. It appears that GPT-3.5 still exhibits noticeable limitations when it comes to the \emph{true/false}, \emph{identify true/false} and \emph{output} MCQs \emph{with code}. While correcting 44 of GPT-3's mistakes it mishandled 20 questions that were answered correctly by the GPT-3 model.

\begin{table}[t]
  \caption{The results of the quantitative evaluation of the differences in performance between GPT-3 and GPT-3.5 across different types of MCQs.}
  \label{tab:gpt3_diff}
  \centering
  \setlength{\tabcolsep}{6pt}
  \begin{tabular}{l|rrrr}
  \toprule
                        
                        &\multicolumn{1}{c}{Mistakes by}&\multicolumn{1}{c}{Fixed by}&\multicolumn{1}{c}{Introduced}&\multicolumn{1}{c}{Fix Ratio}\\
    Question Type       &\multicolumn{1}{c}{GPT-3}&\multicolumn{1}{c}{GPT-3.5}&\multicolumn{1}{c}{by GPT-3.5}& \\
  \midrule
    \bf No Code         & \bf 80          & \bf 52 (65.0\%) & \bf 10    & \bf 5.2   \\
  \midrule
    True/False          & 12              & 11 (91.7\%)     & 4         & 2.8     \\
    Identify True/False & 32              & 20 (62.5\%)     & 5         & 4.0\\
    Finish Statement    & 14              & 9  (64.3\%)     & 1         & 9.0\\
    Other               & 22              & 12 (54.5\%)     & 0         & $\infty$\\
  \midrule
    \bf With Code       & \bf 251         & \bf 140 (55.7\%)& \bf 34    & \bf 4.1     \\
  \midrule
    True/False          & 11              & 4 (36.4\%)      & 6         & 0.67     \\
    Identify True/False & 16              & 5 (31.2\%)      & 4         & 1.25 \\
    Finish Statement    & 17              & 14 (82.4\%)     & 2         & 7.0  \\
    Fill-in             &  8              & 8 (100\%)       & 2         & 4.0  \\      
    Output              & 82              & 35 (42.8\%)     & 10        & 3.5  \\
    Other               & 117             & 74 (63.2\%)     & 10        & 7.4  \\
  \midrule
    \bf Total           & \bf 331         & \bf 192 (58.0\%)& \bf 44    & \bf 4.4   \\
  \bottomrule
  \end{tabular}
\end{table}

Table \ref{tab:gpt4_diff} summarizes the differences between GPT-3.5 and GPT-4. Overall, the GPT-4 model fixed 125 (68.3\%) mistakes committed by GPT-3.5. At the same time, it introduced only 26 novel mistakes. The rate of improvement appears to be higher for the \emph{no code} MCQs (76.3\% of mistakes fixed) compared to the \emph{with code} questions (66.2\%). However, the \emph{with code} questions of the \emph{output} and \emph{other} types are the chief contributors towards the higher performance of GPT-4 as compared to GPT-3.5. For these MCQs, the GPT-4 model corrected 78 questions incorrectly answered by GPT-3.5 while only committing 13 novel mistakes. Overall, it appears that GPT-4 improves over the GPT-3.5 model reliably across all the question types.

\begin{table}[t]
  \caption{The results of the quantitative evaluation of the differences in performance between GPT-3.5 and GPT-4 across different types of MCQs.}
  \label{tab:gpt4_diff}
  \centering
  \setlength{\tabcolsep}{6pt}
  \begin{tabular}{l|rrrr}
  \toprule
                        
                        &Mistakes by      &Fixed by         &Introduced & Fix Ratio \\
    Question Type       &GPT-3.5          &GPT-4            &by GPT-4   & \\
  \midrule
    \bf No Code         & \bf 38          & \bf 29 (76.3\%) & \bf 7     & \bf 4.1  \\
  \midrule
    True/False          & 5               & 4 (80.0\%)      & 1         & 5       \\
    Identify True/False & 17              & 11 (64.7\%)     & 4         & 4.3     \\
    Finish Statement    & 6               & 6  (100\%)      & 0         & $\infty$\\
    Other               & 10              & 8  (80.0\%)     & 2         & 4.0     \\
  \midrule
    \bf With Code       & \bf 145         & \bf 96 (66.2\%) & \bf 19    & \bf 5.1  \\
  \midrule
    True/False          & 13              & 6 (46.2\%)      & 3         & 2.0     \\
    Identify True/False & 15              & 7 (46.7\%)      & 2         & 3.5     \\
    Finish Statement    & 5               & 4 (80.0\%)      & 1         & 4.0     \\
    Fill-in             & 2               & 1 (50.0\%)      & 0         & $\infty$\\      
    Output              & 57              & 40 (70.2\%)     & 7         & 5.7     \\
    Other               & 53              & 38 (71.7\%)     & 6         & 6.3     \\
  \midrule
    \bf Total           & \bf 183         & \bf 125 (68.3\%)& \bf 26    & \bf 4.8   \\
  \bottomrule
  \end{tabular}
\end{table}

\subsection{Qualitative Analysis}
\label{sec:results-qual}

\subsubsection{GPT-3 vs GPT-3.5}
Table \ref{tab:gpt3_qual} defines the 6 prevalent themes identified during the thematic analysis (Section \ref{sec:experimental_design}). The analysis focused on the questions that were answered incorrectly by GPT-3 and, at the same time, handled well by GPT-3.5. In that way, we hoped to uncover traits of questions that appear to be handled much more successfully by the newer model. We also utilize the \emph{Miscellaneous} theme for the MCQs that were not matching any of the prominent themes we identified.

\begin{table}[t]
  \caption{The results of thematic analysis performed on MCQs incorrectly handled by GPT-3 which GPT-3.5 answered correctly.}
  \label{tab:gpt3_qual}
  \centering
  \begin{tabular}{p{4cm}p{7cm}r}
  \toprule
    Theme                   &Definition       &Count \\
  \midrule
  Code Tracing              &MCQs the correct answering of which is contingent on careful tracing of multi-line code snippet shown in the question's stem. &50    \\
  Code in Choices           &MCQs with code snippets in the choices.                &45    \\
  Nuanced String Formatting &MCQs focused on detailed string formatting such as exact number of printed symbols or white space.&27    \\
  Programming Concepts      &MCQs that target understanding of fundamental concepts of Python language.                 &20    \\
  Complex Choices           &More than one choice needed for correct answering, or MCQs asking for the choice that is false.&15    \\
  Arithmetic Expressions    &MCQs centered around solving arithmetic expressions.                 &8    \\
  Miscellaneous             &MCQs that were not assigned to any of the above categories.                 &28    \\
  \bottomrule
  \end{tabular}
\end{table}

The most commonly appearing theme we identified were MCQs with non-trivial \emph{code tracing} required for correct answering. Since the GPT models do not have the direct ability to execute computer code answering such questions may pose a significant challenge. Nevertheless, we observed very noticeable improvement in performance between GPT-3 and GPT-3.5. This suggests that the newer model is capable of handling the MCQs with much more success despite the lack of ability to execute code. The following question is an example MCQ where code tracing is needed:

\begin{quote}
What is the output of the following snippet?

\begin{Verbatim}
def f(x):
    if x == 0:
        return 0
    else:
        return x + f(x - 1)

print(f(3))
\end{Verbatim}

A. 1\\
B. the code is erroneous\\
C. 3\\
\emph{D. 6}
\end{quote}

\noindent Clearly, one would have to follow the execution of the code in order to answer correctly.

The second, almost equally prevalent theme, we identified features questions that have \emph{code in choices}. Such questions may be challenging because they would often involve understanding nuanced differences between the code snippets in order to pick the one which is the correct answer. Apparently, GPT-3.5 has much improved capabilities in answering this type of questions compared to the older GPT-3 model. The following MCQ features multiple code snippets:

\begin{quote}
Which program will produce the following output?

\begin{Verbatim}

Mo Tu We Th Fr Sa Su
\end{Verbatim}

\ 

Program 1

\begin{Verbatim}
import calendar
print(calendar.weekheader(2))    
\end{Verbatim}

\ 

Program 2

\begin{Verbatim}
import calendar
print(calendar.weekheader(3))
\end{Verbatim}

\ 

Program 3

\begin{Verbatim}
import calendar
print(calendar.weekheader())
\end{Verbatim}

\ 

Program 4

\begin{Verbatim}
import calendar
print(calendar.week)
\end{Verbatim}

\ 

\emph{A. Program 1}\\
B. Program 2\\
C. Program 3\\
D. Program 4
\end{quote}

\noindent Answering the above question correctly entails considering the output of each of the four programs. This is likely a significant challenge for the GPT-3 model.

The third theme that clearly emerged from the thematic analysis were MCQs focused on \emph{nuanced string formatting}. Such questions revolve around details of the code snippet output such as exact number of printed symbols or white space. GPT-3.5 appears to be much more sensitive towards such details and, hence, handles the MCQs much more successfully. The following question asks about the number of asterisks (\verb|*|) printed to the terminal:

\begin{quote}
How many stars (*) will the following snippet send to console?

\begin{Verbatim}
i = 0
while i < i + 2:
    i += 1
    print('*')
else:
    print('*')
\end{Verbatim}

A. one\\
B. zero\\
C. two\\
\emph{D. the snippet will enter an infinite loop, printing one star per line}
\end{quote}

\noindent This type of question may also be challenging because correct answering requires meticulous code tracing. Hence, this theme might be somewhat related to the \emph{code tracing} theme described above.

The next theme that emerged is somewhat surprising as it describes relatively simple questions that target understanding of fundamental concepts of Python language (i.e., \emph{programming concepts}). Such MCQs typically present the least challenge to students as they mostly operate on the lower levels of Bloom's taxonomy (remember, understand). The following question tests the conceptual understanding of the exception handling in Python:

\begin{quote}
If there are more than one except: branches after the try: clause, we can say that:

A. exactly one except: block will be executed\\
\emph{B. not more than one except: block will be executed}\\
C. one or more except: blocks will be executed\\
D. none of the except: blocks will be executed
\end{quote}

\noindent While it is quite counter-intuitive that GPT-3 struggles with questions such as this one GPT-3.5 appears to handle them much more successfully.

The next theme that we identified through the thematic analysis describes MCQs with \emph{complex choices}. We deemed a questions as having complex choices if it either required more than one choice to be selected as the correct answer or if a false (incorrect) statement needs to be identified. Such questions likely pose a challenge to GPT-3. The more capable GPT-3.5 model is able to deal with such complexities much more successfully. The below question is an example of MCQ where a false statement needs to be identified as the correct answer:

\begin{quote}
Which of the following statements is false?

A. pip is the most popular package-management system for Python\\
B. pip connects with the Python Package Index to help developers install and manage software packages.\\
C. pip allows you to install the specific version of a package from the command line\\
\emph{D. pip is one of the most widely used repositories for finding and publishing Python packages}
\end{quote}

\noindent It is likely more challenging for the LLM to associate the false choice with the question as the correct answer. Hence, a more powerful model is needed to handle such questions successfully.

It is quite well-known that GPT models struggle with often even simple math. Hence, it was not that surprising that MCQs organized around \emph{arithmetical expressions} posed a serious challenge to the GPT-3 model. The below question is an example of MCQ with an arithmetic expression:

\begin{quote}
Evaluate the following expression and determine whether it is True or False.

\begin{Verbatim}
2 + 2 != 2 * 2    
\end{Verbatim}

\emph{A. False}\\
B. True
\end{quote}

\subsubsection{GPT-3.5 vs GPT-4}
Table \ref{tab:gpt4_qual} describes the 6 prominent themes we identified, performing the thematic analysis as described in Section \ref{sec:experimental_design}, among the MCQs that were not answered correctly by GPT-3.5 but that were handled properly by GPT-4. Hence, the themes provide detailed insight into systematic phenomena where GPT-4's performance appears to be noticeably improved when compared to GPT-3.5. We also list the \emph{Miscellaneous} category that was assigned to the MCQs that we could not associate with any prominent pattern.

\begin{table}[t]
  \caption{The results of thematic analysis performed on MCQs incorrectly handled by GPT-3.5 which GPT-4 answered correctly.}
  \label{tab:gpt4_qual}
  \centering
  \begin{tabular}{p{4cm}p{7cm}r}
  \toprule
    Theme                   &Definition       &Count \\
  \midrule
  Specific Code Constructs  &MCQs involving at least one of the following code constructs: exception handling, variable re-assignment, indexing.  &27    \\
  Extensive Code            &MCQs with larger blocks of code ($>10$ lines) or multiple code snippets (e.g., a block of code in the stem as well as the choices).                &26    \\
  Nuanced String Formatting &MCQs focused on detailed string formatting such as exact number of printed symbols or white space.&21    \\
  Programming Concepts      &MCQs that target understanding of fundamental concepts of Python language.                 &16    \\
  Arithmetic Expressions    &MCQs centered around solving arithmetic expressions.                 &11    \\
  Complex Choices           &More than one choice needed for correct answering, or MCQs asking for the choice that is false.&10    \\
  Miscellaneous             &MCQs that were not assigned to any of the above categories.                 &14    \\
  \bottomrule
  \end{tabular}
\end{table}

The most prominent theme we identified were MCQs that involved one or more of the following \emph{specific code constructs}: exception handling, variable re-assignment or indexing. It appears that the GPT-3.5 model systematically struggled with these constructs whereas GPT-4 handles them relatively well. The following question is an example of the variable re-assignment:

\begin{quote}
What is the output of the following snippet if the user enters two lines containing `2` and `4` respectively?

\begin{Verbatim}
x = int(input())
y = int(input())

x = x // y
y = y // x

print(y)
\end{Verbatim}

\emph{A. the code will cause a runtime error}\\
B. 2.0\\
C. 8.0\\
D. 4.0
\end{quote}

\noindent Observe that variables \verb|x| and \verb|y| are assigned values multiple times throughout the execution of the program. We hypothesize that this type of question is related to the \emph{code tracing} theme identified in the previous thematic analysis where GPT-3 was compared to GPT-3.5. While GPT-3.5 clearly improved in this regard upon the performance of GPT-3 it appears that certain constructs, such as variable re-assignment, exception handling or indexing, remained challenging. While these may still be challenging for GPT-3.5 the GPT-4 model is much more successful in answering these types of MCQs.

Another common theme were questions that contained larger blocks of code or multiple code snippets (i.e., \emph{extensive code} element). These appear to be handled much more successfully by the more recent GPT-4 model. This theme is likely related to the \emph{code in choices} theme identified in the preceding analysis, comparing GPT-3 to GPT-3.5. Despite the significant improvement brought about by GPT-3.5 it appears that larger or multiple code blocks still remained somewhat challenging for the model, leaving space for further improvements by GPT-4.

The remaining four themes that emerged, i.e., the \emph{nuanced string formatting}, \emph{programming concepts}, \emph{arithmetic expressions} and \emph{complex choices} are the same as in the preceding analysis performed for the GPT-3 and GPT 3.5 models. Hence, while we observed sizeable improvement between the two models these challenges somewhat remained. GPT-4 then offered further improvements.

\section{Implications for Teaching Practice}
This research reinforces and builds upon prior investigations \cite{savelka2023thrilled}, indicating that programming educators should anticipate a future where learners can effortlessly use readily available technology to earn passing marks from current programming knowledge and skills evaluations. While this development has been apparent from the growing body of prior work \cite{Savelka2023,10.1145/3511861.3511863,finnie2023my,denny2023conversing,piccolo2023many} this paper provides detailed insight into the qualitative differences among the studied GPT models.

Given this backdrop, educators might consider putting a deeper emphasis on learning as opposed to assessment. This involves accentuating the overall learning journey and skill acquisition over merely prepping students for tests. Learners should be encouraged towards personal growth, rather than just producing the correct answers. The pivotal role of academic integrity and ethical standards within the academic setting must be underlined. The aim should be to foster an environment where originality and individual diligence are treasured. Traditional modes of assessment, such as MCQs, may become less relevant, giving way to novel evaluation methods that necessitate on-the-spot demonstration of understanding.

While discerning GPT models' shortcomings in tackling MCQs might seem like an attractive strategy for test design, we contend that it is potentially short-lived. Based on the rapid advancements charted in this research, it is plausible that current gaps will soon be bridged, rendering such ``GPT-centric'' tests ineffective. Rather than crafting ``GPT-resistant'' exams, a more promising trajectory might be the creation of evaluations centered on higher cognitive capabilities such as analytical reasoning, innovative problem-solving, and inventive thinking---areas where GPT models still grapple.

\section{Limitations and Threats to Validity}
Although, the results of our study provide important insights into the evolving capabilities of the GPT models in answering MCQs from introductory and intermediate Python classes at the higher education level, limitations in several areas must be acknowledged. While Python is a widely used programming language, it is merely one among a plethora of languages. Although, GPT models are adept at processing various programming languages, our results might not necessarily extrapolate to other languages characterized by distinct structural, syntactical, and conventional nuances. Furthermore, our study's scope was confined to English-language MCQs.

The exact dataset underpinning the OpenAI's models' training remains unknown. If the MCQs were already part of the training data, our tests would not demonstrate the models' innate proficiency in tackling assessments. Instead, they would underscore their retention prowess. Although we are fairly confident that our chosen assessments had not been exposed during model training-—-given their absence in known public datasets--—it is an inherent limitation that researchers must recognize when evaluating OpenAI's GPT models. 

\section{Conclusions and Future Work}
\label{sec:conclusions}
We conducted a comprehensive quantitative and qualitative evaluation of three GPT models using a robust collection of 530 MCQs, with many incorporating code excerpts, taken from three distinct Python programming courses. The research underscored the models' evolving proficiency in passing MCQ evaluations, illustrated through an in-depth thematic analysis. A primary takeaway is the looming threat of students leaning excessively on GPT models for programming evaluations---a sentiment resonating with  \cite{becker2022programming,prather2023robots}. Given this context, it's imperative to devise methods to tackle this escalating issue, ensuring the continued integrity of programming assessment.

Though our exploration into the GPT models' capabilities across a range of MCQs provided many insights, it offers ample space for future work. Therefore, we propose several potential paths for continued research: (i)~delving deeper into the implications of prompt adjustments; (ii) gauging the prowess of GPT models in other areas, such as competitive math; and (iv) investigating the prospects of seamlessly incorporating GPT-driven resources, such as ChatGPT or Copilot, into programming education.

\bibliographystyle{splncs04}
\bibliography{refs}
%








\end{document}